# A Kubernetes Controller for Managing the Availability of Elastic Microservice Based Stateful Applications[*]

Leila Abdollahi Vayghan
Engineering and Computer Science
Concordia University
Montreal, Canada
l_abdoll@encs.concordia.ca

Mohamed Aymen Saied
Computer Science and Software Engineering Department
Laval University
Quebec, Canada
mohamed-aymen.saied@ift.ulaval.ca

Maria Toeroe
Ericsson Inc.
Montreal, Canada
maria.toeroe@ericsson.com

Ferhat Khendek
Engineering and Computer Science
Concordia University
Montreal, Canada
ferhat.khendek@concordia.ca

*Abstract*— The architectural style of microservices has been gaining popularity in recent years. In this architectural style, small and loosely coupled modules are deployed and scaled independently to compose cloud-native applications. Carrier-grade service providers are migrating their legacy applications to a microservice based architecture running on Kubernetes which is an open source platform for orchestrating containerized microservice based applications. However, in this migration, service availability remains a concern. Service availability is measured as the percentage of time the service is provisioned. High Availability (HA) is achieved when the service is available at least 99.999% of the time. In this paper, we identify possible architectures for deploying stateful microservice based applications with Kubernetes and evaluate Kubernetes from the perspective of availability it provides for its managed applications. The results of our experiments show that the repair actions of Kubernetes cannot satisfy HA requirements, and in some cases cannot guarantee service recovery. Therefore, we propose an HA State Controller which integrates with Kubernetes and allows for application state replication and automatic service redirection to the healthy microservice instances by enabling service recovery in addition to the repair actions of Kubernetes. Based on experiments we evaluate our solution and compare the different architectures from the perspective of availability and scaling overhead. The results of our investigations show that our solution can improve the recovery time of stateful microservice based applications by 50%.

*Keywords— Microservices; Containers; Kubernetes; Failure; Availability; Elasticity*

## I. INTRODUCTION

With the adoption of cloud computing [1], the microservices architectural style [2] has drawn a substantial amount of attention in the software engineering community. The microservice based architecture tackles the challenges of building cloud-native applications that leverage the opportunities given by the cloud infrastructure [3]. Microservices [4] are a realization of the service-oriented architectural style of building software composed of small services that can be deployed and scaled independently [5]. Each microservice has a separate business functionality, runs as a separate process, and communicates through lightweight mechanisms [2].

The fine granularity of this architectural style makes the scaling flexible and efficient as each microservice can evolve at its own pace. Moreover, compared to monolithic applications, the small microservices can restart faster in case of failure recovery or at the time of their upgrade. Microservices are loosely coupled and the failure of one microservice should not affect other microservices of the system. Because of these characteristics, adopting the architectural style of microservices can improve the service availability of applications [2]. Service availability is a non-functional requirement defined as the percentage of time a service is provisioned [6]. High Availability (HA) is achieved when the system is available at least 99.999% of the time. Therefore, the total downtime allowed in one year for highly available systems is around 5 minutes [7].

To leverage the benefits of microservice based architectures, one needs to use technologies aligned with the characteristics of this architectural style. Containerization is a technology which enables virtualization at the operating system level [8]. Containers are lightweight and portable and therefore, they are suitable for building microservices. Docker [9] is the leading container platform which packages the application code and its dependencies together to ship them as a single container image. Containers are running instances of container images. Even containers running on the same machine are isolated and are not aware of each other. Thus, there is a need for an orchestration platform to manage the deployment of containers. Kubernetes [10] is an open-source container orchestration platform which enables the automated deployment, scaling, and management of containerized applications. Kubernetes, through its monitoring and auto-healing mechanisms, alleviates the complexity of implementing application resiliency. Kubernetes has become the leading orchestration platform for containerized microservice based applications.

Organizations migrate their legacy applications to cloud-native architectures by adopting the architectural style of microservices [11]. These microservice based applications are containerized and orchestrated by Kubernetes predominantly.

---

[*] This paper is an extended version of [13].

The characteristics of microservices and containers – such as being small and lightweight – naturally contribute to improving service availability [12]. Kubernetes heals its managed microservices [10] by restarting the failed containers and replacing or rescheduling them when their hosts fail. Moreover, to ensure access only to healthy containers, Kubernetes does not advertise the unhealthy containers until they are ready again. Nevertheless, these measures may not be sufficient for carrier-grade service providers and service availability as an important non-functional requirement still remains a concern for them.

Redundancy is the most adopted mechanism for enabling HA [6] and Kubernetes automatically replicates microservice instances considering them independent. However, such replication of microservice instances can only improve the service availability of stateless microservice based applications. The reason is that only stateless microservices have interchangeable instances and therefore can be easily replaced. The same does not hold for stateful microservices, whose current operations depend on the result of their previous operations.

With stateful microservice based applications, each microservice instance has its own state, which need to be known to other microservice instances to be able to take over when it fails. Therefore, replicating a stateful microservice instance requires replicating its state to maintain its service availability with continuity. To address this issue, stateful microservice based applications are often deployed as microservice instances that store their states in external databases making them virtually stateless. The assumption is that the database takes care of the data replication. Unfortunately, this replication does not necessarily improve service availability. Indeed, although the state is replicated and available in the database for other instances, the microservice instances are not aware of each other and each other's failure. Hence, they cannot automatically resume the service of each other if failure happens. Kubernetes also provides some support for the deployment of stateful microservice based applications. In this paper, we discuss the reference architectures offered by Kubernetes for managing stateful microservices; and we discuss their availability challenges. In a previous work [13], to address these challenges we proposed an HA State Controller that improves the availability of stateful microservice based applications deployed with Kubernetes. This HA State Controller allows for the automatic service redirection to the healthy microservice instances through the management of secondary labels. However, the elasticity of such microservice based applications was not considered. Our solution in [13] only allowed for one active and one standby microservice instances.

This paper extends our previous work [13] as follows:

- We enhance the HA State Controller to manage the availability of an application as it scales in and out by allowing multiple active and standby assignments. In other words we enhance the HA State Controller in [13] to work with multiple active and standby instances, which may dynamically change over time due to workload change.

- We evaluate the enhanced HA State Controller from the perspectives of availability and scaling overhead through a series of experiments. The goal of our experiments is to answer the following research questions (RQ).
  - **RQ1**: What is the impact of the HA State Controller on the provided availability?
  - **RQ2**: What is the impact of scaling during failover on the availability that the HA State Controller can provide for its managed microservices?
  - **RQ3**: What is the overhead of the HA State Controller at scaling?
  - **RQ4**: What is the impact of simultaneous failures of active pods on the outage of each failed pod?

The rest of the paper is organized as follows. In Section II, we provide some background information on Kubernetes. In Section III, we discuss the Kubernetes architectures for deploying stateful microservice based applications and their challenges with respect to availability. To address these challenges, we propose a solution in Section IV and evaluate our proposal in Section V through a set of experiments. We discuss the related work in Section VI, followed by a conclusion in Section VII.

II. BACKGROUND

Kubernetes is the leading orchestration platform for containerized applications.

A Kubernetes's cluster can be composed of virtual or physical machines and it follows the master-slave architecture. The processes required to bring the cluster to a desired state run on the master node. To ensure the HA of the cluster management, the master node and its processes should be replicated [14].

Kubernetes deploys and manages containerized microservices in the form of pods [10]. A pod is the smallest unit whose lifecycle is managed by Kubernetes. Multiple containers can be grouped into a single pod. They will all get the same IP address, which is the virtual IP address of the pod. Containers within a pod may be accessed separately through different ports. Considering that one of the goals of the microservice based architecture is to make the building blocks of an application loosely coupled, it is recommended to include only one container in each pod [10].

Controllers are entities that manage the pods' lifecycle and responsible for creating and maintaining the required number of pods according to the specification given in the pod template in the controller's resource specfication. Kubernetes has different types of controllers each of which is used for a different purpose. For example, Deployment controllers are mainly used for managing stateless application pods while StatefulSet controllers manage stateful application pods [10].

Controllers delete and reschedule pods dynamically based on changes in the cluster. Therefore, the IP addresses of pods change often and are not reliable for communication. Instead Kubernetes offers customizable labels assigned to pods to select them, e.g., for communication based on these labels. From Kubernetes perspective these labels stay the same when pods are rescheduled. Kubernetes also defines an abstraction called Service, which has a static virtual IP address. Services select pods as their communication endpoints based on their labels. All requests received at the IP address of a service are load balanced among the service endpoints in a random or round-robin manner [10].

As mentioned above, in Kubernetes stateful application pods are deployed by StatefulSet controllers. To cater for the statefulness, StatefulSet pods are provided with storage for their state data. Kubernetes abstracts the details of storage provisioning by two API resources: Persistent Volume (PV) and Persistent Volume Claim (PVC). A PV is a piece of storage in the cluster whose lifecycle is independent of the pods using it. PVs can be provisioned dynamically or statically. A PVC, on the other hand, is a request for storage made by a pod that binds the pod to a specific PV. Binding means that if the pod is restarted or rescheduled with the same identity, its PVC will be assigned to the same PV [10]. Moreover, in the event where the StatefulSet controller receives a request to scale-in the application, it will only remove the required number of pods and will keep the PVCs. However, the state data stored on the PVCs will not be reachable by clients until their corresponding pods are recreated. For stateful applications, the assumption is that the application state data are stored on a persistent storage represented by the PV, which is outside of the Kubernetes cluster. Meaning that, Kubernetes is not in charge of managing the PVs and it only consumes them.

## III. STATEFUL MICROSERVICE BASED ARCHITECTURES AND AVAILABILITY MANAGEMENT WITH KUBERNETES

In this section, we describe the Kubernetes architectures offered for deploying pods of stateful microservice based applications and discuss their challenges with respect to availability.

### A. Architectures for Deploying Stateful Applications with Kubernetes

For stateful microservice based applications, Kubernetes provides different solutions. Kubernetes' primary controller for deploying stateful applications is the StatefulSet controller. A StatefulSet controller specification includes a PVC template which describes the characteristics of the PVs such as capacity, access mode, etc., that a pod of the StatefulSet can be bound to. For each pod created by a StatefulSet controller, a PVC is created based on the PVC template that binds the pod to a PV, which matches the criteria described in the PVC template. As a result, each pod has a separate PV where it can store its state data. A PV is accessible only by the pod it is bound to. This means, that a pod cannot access other pods' PVs and state data. Therefore, a mechanism such as sticky sessions is needed to ensure that a client is always served by the same server (i.e., same pod). The reason is that a client's request should always be sent to a pod who has the client's state data.

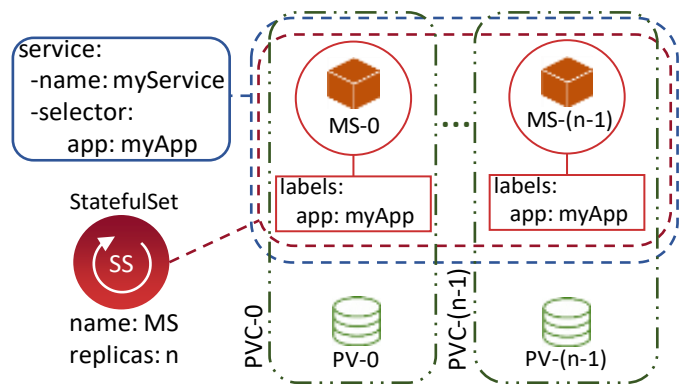

Fig. 1. Deploying stateful applications with StatefulSet controllers.

Fig. 1 shows the architecture for deploying stateful applications using a StatefulSet controller.

As shown in Fig. 1, the names of StatefulSet pods are composed of the name of the StatefulSet controller ("MS") and an ordinal index (MS-0, MS-1… MS-(n-1)). One of the differences between StatefulSet pods compared to pods managed by other controllers is that they are created and deleted sequentially. That is, MS-1 will be created only when MS-0 is running and ready. Another difference is that StatefulSet pods have persistent identities. Meaning that if MS-0, which stores its state data in PV0, fails, the StatefulSet controller will restart the pod with the same identity. Therefore, the new incarnation of MS-0 will be bound to PV0 again, and it will have access to its state data stored by its previous incarnations.

Although StatefulSet controllers are the recommended controllers for deploying stateful application pods, one can use Deployment controllers for this type of application as well. Similar to StatefulSets, the stateful Deployment pods can store their state data in a PV. However, with Deployment controllers, all pods have to share the same PV. Accordingly there is no PVC template in a Deployment controller specification. Instead, one PVC is created before deploying the application and it will be used by all pods of the application once they are deployed by the Deployment controller. Fig. 2 shows the architecture for deploying stateful applications using a Deployment controller. In this architecture, the PV is shared between all pods. Therefore, all pods can have access to all state data and could serve any client.

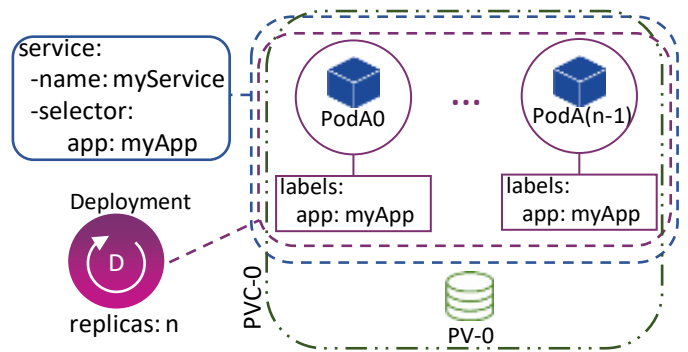

Fig. 2. Deploying stateful applications with Deployment controllers.

*B. Availability Challenges*

Kubernetes provides an auto-healing mechanism for the applications it manages through restarting their failed containers/pods on the same host, or rescheduling them on another when their host fails. Although these repair actions can improve the availability of the applications deployed with Kubernetes, state replication remains the most important feature to achieve high availability (HA).

As discussed before, Kubernetes' controllers such as StatefulSet and Deployment are able to maintain multiple pod replicas. However, from an availability perspective, only stateless applications benefit from replicating their pods. The reason is that that the mechanism provided for replicating the state of the pods is not sufficient. The availability of stateful applications is not improved by increasing the number of pod replicas. For stateful applications still the only mechanism for maintaining availability with Kubernetes is through the repair of the failed pods. Let us explain our view through an example.

If the application is deployed by a StatefulSet controller as shown in Fig. 1, if one pod fails, the other pods cannot resume the service of the failed pod because: (1) the state data for each pod are stored separately and other pods do not have access to them. (2) the pods are isolated and are not aware of each other's failure. Therefore, we can only rely on the failed pod being restarted with the same identity so it can restore from its own PV the last stored state before the failure. This means that the service can be recovered with continuity, but the clients need to wait for the failed pod to be restarted, which may be too slow for some applications.

Moreover, in certain failure scenarios, the failed pod will not be repaired by Kubernetes. For example, with the architecture in Fig. 1, if the service outage is due to node shutdown, the pod will not be restarted and the service will not be recovered unless the node rejoins the cluster. The reason is that Kubernetes cannot differentiate between node failure and network partitioning, and to avoid having multiple pods running on different nodes with the same identity, it will not automatically create a pod to replace the one on the unresponsive node. Unlike StatefulSet controllers, pods deployed by Deployment controllers (Fig. 2) are automatically rescheduled on other nodes when their hosts fail. However, this would not mean that the failed pod's service is recovered. The reason is that Deployment controller pods do not have sticky identities and after restart, they will have a new identifier and they will not be aware of the identity or the location where the failed pod's state data is stored. Thus, we cannot rely on the restart procedure for recovering the stored service state.

In a previous work [13], we evaluated the level of availability Kubernetes can provide for stateful applications solely through its repair actions. In our experiments in the scenarios where service outage was due to application container failure, the total service outage was 2.159 seconds on average. This meant that for our application to meet the high availability requirements, no more than 146 application container failures could be tolerated in one year. We also conducted experiments with the scenario where the service outage was due to node reboot. In this scenario, the node became responsive again and rejoined the cluster and therefore, the service was recovered. However, the service recovery depended on the node start-up time, which in our experiments was measured 126.4 seconds on average, which resulted in a total outage of 164.507 seconds. This meant that to meet high availability requirements, only one failure due to node reboot could be tolerated in one year.

These experiments showed that relying on the repair actions of Kubernetes was not enough for satisfying high availability requirements and additional mechanisms were needed for Kubernetes to decouple service recovery from the repair of the failed pod.

In the next section, we introduce our solution that integrates with Kubernetes and improves the availability by recovering the service before the failed pod is repaired by Kubernetes.

IV. A STATE CONTROLLER ENHANCED WITH ELASTICITY

As mentioned earlier, the main challenge with respect to availability for the applications deployed with Kubernetes is that in case of failure, the failed pod should be repaired for the service to be recovered. It is possible to address this issue by keeping a redundant pod (i.e., a standby pod) which has the state of the failed pod that was providing the service (i.e., active pod), and therefore it can take over in providing the service. In addition to the active pod replicating its state to the standby pod, it is necessary for the standby pod to be notified when its active pod has failed. Moreover, elasticity also needs to be considered. That is, multiple active pods providing a service should be possible whose states need to be protected by standby pods. It should be possible to scale out/in the pods while maintaining service availability.

We address these issues by a solution which integrates the concept of HA states (i.e., active and standby) with Kubernetes and improves the availability of stateful microservice based applications by recovering the service before the failed pod has been repaired. In this solution, a HA State Controller (SC) component is integrated with Kubernetes. It communicates with the Kubernetes API server and monitors the cluster events and reacts to them. The proposed SC assigns an HA state to running pods and provides a mechanism for the active pod to replicate its state data to its standby pod. The SC detects the scale-out and scale-in events and reacts to them by assigning to and removing from pods their HA states. The State Controller can have more than one state replication service created automatically. The State Controller holds pairs of pods as active and standby and identifies a pair by adding a "peer" label to the standby pod, which gets its corresponding active pod's name. The SC also reacts to pod failures. In case of failure of an active pod it notifies its corresponding standby pod for initiating the failover process.

Fig. 3 provides an overall picture of the State Controller behavior. The first part of this diagram is about HA state assignment and pod labelling while the second half is about the State Controller reactions to the events. In addition to this high level behavior of the state controller we provide a lower level algorithm (Algorithm 1). Hereafter, we summarize the steps in this high level diagram and put them in relation with the lower level algorithm (Algorithm 1), showing for each step how it is implemented and the corresponding instructions in Algorithm 1 using labels.

1. The State Controller sorts running pods based on their creation time (L30)

2. As long as there are pods without HA state and peer labels (L31:L39):

    a. It picks the oldest two pods (L33)

    b. It assigns HA state and peer labels to both pods and removes them from the sorted list (L34:L38)

3. The State Controller now watches the events of the Kubernetes API server (L4:L27)

    a. If the event corresponds to the service state of a pod changing to "not ready" then this is a failure event (L7:L19)

        i. If the failed pod had the active HA state (L9:L15)

            1. The State Controller assigns active HAState to the standby pod which was the peer of the failed active pod. The new active pod becomes the endpoint of the application service and restores the last state from its storage area in the PV and resumes the service.

            2. The State Controller assigns standby HAState to the failed pod and deletes the state replication service of the failed active pod

            3. The State Controller creates the replication service for the new active pod

        ii. If the failed pod had the standby HAState, the State Controller ensures that the failed pod is assigned the standby HAState after it is repaired (L16:L19)

    b. If the event corresponds to a scaling event (L20:L26)

        i. In case of a scale out, then go to step 1 (L22)

        ii. In case of a scale in, the State Controller deletes the state replication service for a deleted active-standby pair (L25)

In the following subsections, we elaborate on these different responsibilities of the State Controller (i.e. HA state assignment, state replication, handling elasticity, handling failures).

The proposed SC can be integrated with Kubernetes without any change to the Kubernetes source code. It can be integrated with StatefulSet controllers (see Fig. 4) as well as with Deployment controllers (see Fig. 5). The difference between these two architectures is in the way the pods of each controller store their state data. As discussed in Section III, with StatefulSet controllers, each pod has a separate persistent volume (Fig. 4) while with Deployment controllers all pods share the same persistent volume (Fig. 5). Therefore, in the latter, each pod creates a storage area for itself when it is deployed in order to separate its state from other pods.

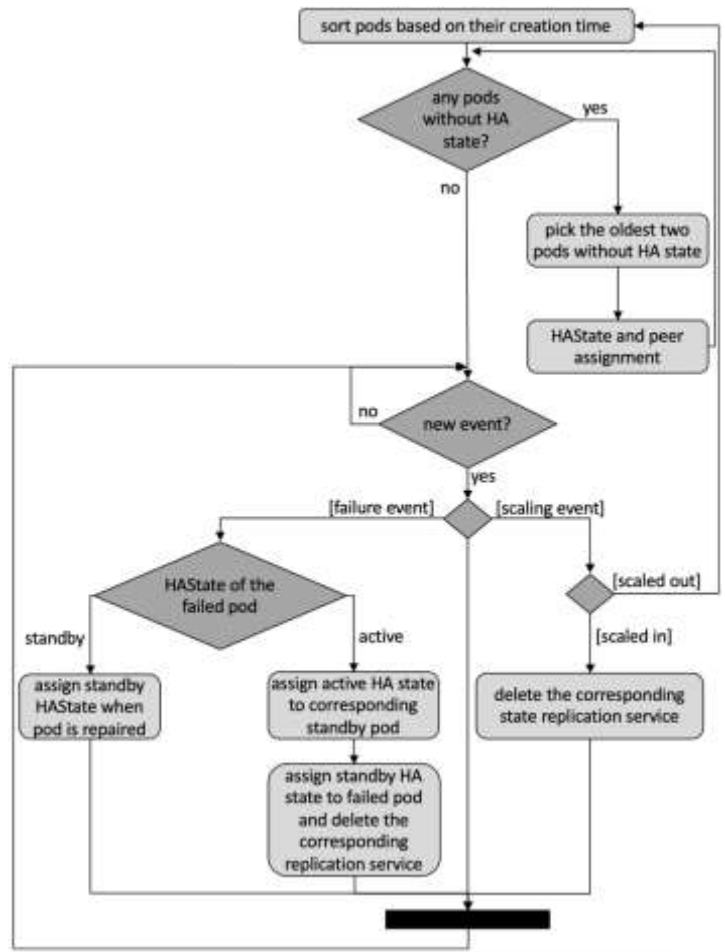

Fig. 3. Activity diagram of the HA State Controller.

### A. HA State Assignment

One role of the SC is to assign an HA state to pods through the management of secondary labels and environment variables. That is, determining whether a pod should be active or standby.

Algorithm 1. HA State Controller algorithm

```
1.  HAStateController (runningPodsList){
2.  assignsHAstateAndPeerLabels(runningPodsList)
3.
4.  WHILE (true)
5.        //watches the events of the Kubernetes API server
6.        event ← KubernetesAPIserver.getEvent()
7.         IF(event.type==podFailure) THEN
8.                   handledPod = event.getFailedPod()
9.                   IF(handledPod.getHastate()== active) THEN
10.                  newActivePod= handledPod.getPeer()
11.                  newActivePod.setHastate(active)
12.                  // The failed pod will be restarted by kubernetes
13.                  handledPod.setHAstate(standby)
14.                  deletesStateReplicationService(handledPod, newActivePod)
15.                  createsStateReplicationService(newActivePod, handledPod)
16.                ELSE IF(handledPod.getHAState()== standby) THEN
17.                  // The failed pod will be restarted by kubernetes
18.                handledPod.setHAState(standby)
19.                End IF
20.       ELSE IF (event.type== scaleOut) THEN
21.                 scaledOutPodList←event.getListOfAddedPods()
22.                 assignsHAStateAndPeerLabels(scaledPodList)
23.       ELSE IF (event.type== scaleIn) THEN
24.                scaledInPodList← event.getListOfDeletedPods()
25.                deletesStateReplicationService(scaledInPodList[1],scaledInPodList[2]))
26.         End IF
27. END WHILE
28. }
29. assignsHAstateAndPeerLabels(podsList){
30.      podList←sortPodsBasedOnCreationTime (podsList)
31.      WHILE ( podList.HasNext()) do
32.           // we made the assumption that the number of pods is even
33.            (podA,podB)← podList.getNextTwoPod()
34.            podList.remove(podA)
35.            podList.remove(podB)
36.            assignsHAstate(podA,podB)
37.            assignsPeerLabels(podA,podB)
38.            createsStateReplicationService(podA,podB)
39.     END WHILE
40. }
```

In order to do so, the SC communicates with the API server and gets the list of running pods and sorts them based on their creation time. Out of this ordered list of pods, the SC selects pairs of pods and for each pair, it assigns a secondary label called the "HAState" label with the value of active to one pod and standby to the other pod. The assumption is that an even number of pods are deployed. In this solution, the service that exposes the application (i.e., application service) redirects the incoming requests to pods that have the HAState label with the value of active. Therefore, when the SC assigns active HA state to pods, it adds these pods to the endpoints list of the application service. In addition, the SC identifies pod-pairs by adding a "peer" label to pods. The standby pod's peer label is set to its corresponding active pod's name and the active pod's "peer" label gets its corresponding standby pod's name.

In addition to the labels, the SC also creates and populates an environment variable – the HAState variable – within each pod to make the pod aware of its HA state. A process is included in the container image of all pods, which we call the entrypoint process, periodically checks the "HAState" variable and makes decisions according to the changes that the SC makes to the "HAState" variable. For example, if the "HAState" variable changes from standby to active, it will call the service resume process.

## B. State Replication

Besides assigning HA states to pods, the SC implements a mechanism for the active pod to replicate its state data to its standby pod. The SC enables the state replication by automatically creating a Kubernetes service for each pair of active-standby that exposes the standby pod to the active pod. The state replication service redirects the incoming requests to a pod that has the "HAState" label with the value of standby and the "peer" label with the value of the active pod's name. The SC names the state replication service for each pair as "replicate-{the active pod's name}". Therefore, the active pod does not need to know the IP address and can discover the service based on this naming pattern. For example, an active pod named PodA0 sends its state data by an HTTP request to a state replication service named "replicate-PodA0". This is done in the pod's checkpointing process, which periodically also saves the state data in the storage. This way, the standby pod will have the state of the active pod. Although the checkpointing process for every application can be implemented in a different way, the state replication service is application agnostic as long as the application is able to store and transfer its state data through HTTP requests.

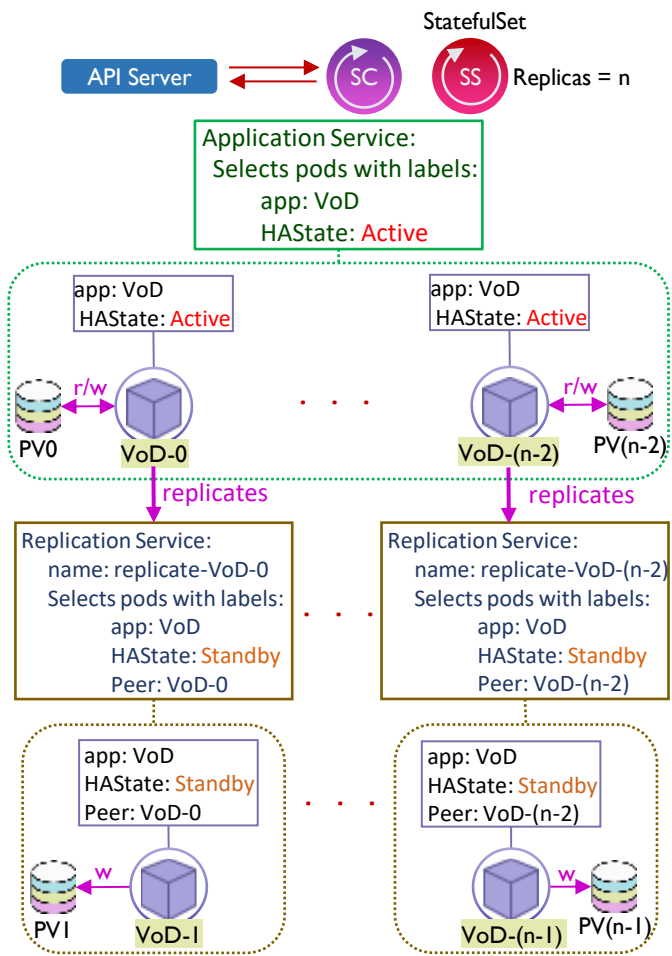

Fig. 4. Integrating the State Controller with StatefulSet controllers.

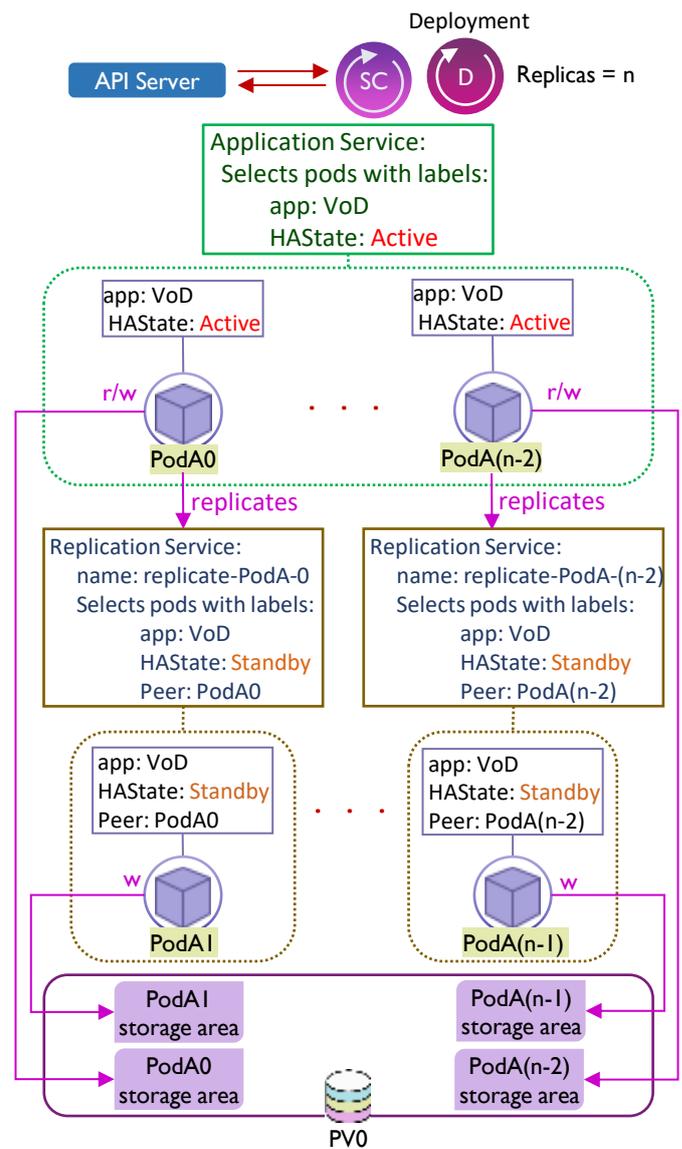

Fig. 5. Integrating the HA State Controller with Deployment controllers.

*C. Handling Elasticity*

For microservice based applications, it is common for the number of microservice instances to increase or decrease frequently. Therefore, we designed the SC in such a way that it can handle multiple active and standby assignments, it detects the scale-out and scale-in events, and reacts to the addition and deletion of pods. The assumption is that the pods are added or deleted in pairs. For the scale-out scenario, for every pair of pods that are deployed, the SC assigns active and standby HA state to pods and creates a state replication service. The SC guarantees service protection for scale-in scenario only if the application is deployed by a Statefulset controller. The reason is that unlike Deployment controllers, when a Statefulset controller receives a scale-in request, it deletes the requested number of pods in the reverse order that they were created. The SC reacts to the scale-in event by removing the pairs from its list and deletes their corresponding state replication services.

*D. Handling Failures*

An important task of the SC is to detect the failure of active pods and inform their corresponding standby pods to resume the service, which was provided by the failed active pod. When a pod fails, its service state changes into "not ready" and this change is recorded as an event by the API server. To detect the failure of pods the SC monitors the API server events. If the failed pod had the standby HA state, the SC will wait for the pod to be repaired and will assign the standby HA state to the repaired pod again. If the failed pod had the active HA state, the procedure will be different. In this case, the SC will change the "HAState" label and variable of the standby pod corresponding to the failed active pod from standby to active. As mentioned before, when the "HAState" variable changes from standby to active, the entrypoint process will call the service resume process. Since the new active pod has the last saved state of the failed active pod, it is able to resume the service from that point as soon as it is added to the endpoint list of the application service due to the change the "HAState" label. The SC also removes the corresponding state replication service. The SC will assign the standby HA state to the failed pod after it is repaired. Moreover, a new state replication service will be created to which the newly active pod replicates its state data.

## V. AVAILABILITY AND PERFORMANCE EVALUATION

In this paper, we evaluate through a set of experiments the achievable availability as well as the scaling overhead of integrating the SC with Kubernetes.

*A. Experiments' Settings*

The setting for these experiments is a Kubernetes cluster in a private cloud which is composed of eight worker nodes and one master node running on the OpenStack cloud. Ubuntu 16.04 is the OS running on all nodes. Kubernetes 1.12.1 runs on all VMs and the container engine is Docker 17.09. Network Time Protocol (NTP) [15] is used for time synchronization between the nodes. The application deployed is a stateful Video on Demand (VoD) application, where each client can request a video to be streamed. The same pod template is used for all experiments that has one container image in which the VideoLan Client (VLC) [16] is installed as the video streaming application. To ensure service continuity, the container image has a checkpointing process which, whenever the pod receives a request from a client to stream a video, starts to checkpoint the elapsed time of the video to the location where its PV is mounted. The streaming position, which is the state data in this case, is stored for each client separately.

*B. Metrics*

In our experiments, we measure the following metrics:

**Availability metrics:** The metrics we measure to evaluate the availability in our experiments are composed of reaction time, repair time, recovery time, and total outage time. The *reaction time* is measured as the time between the failure event and the first reaction of Kubernetes that reflects that the failure event was detected. The *repair time* is the time between the first reaction of Kubernetes and the repair of the pod failed due to the failure event. The *recovery time* is the time between the first reaction of Kubernetes to the failure event and when the service is available again. The total *outage time*, that is, the duration for which the service was not available, is the sum of the reaction time and the recovery time. These metrics and their relations are shown in Fig. 6. In Fig. 6 (a) the sequence of events is for Kubernetes without the SC whereas in Fig.6 (b), the sequence is for the architectures where the SC is integrated. As it is depicted in Fig. 6, the SC reduces the outage time by recovering the service before the failed unit is repaired.

**Scaling time:** The delay from the moment of sending the scaling request until the last pod is deployed and ready (or deleted, in case of scale in) in reaction to the scaling request.

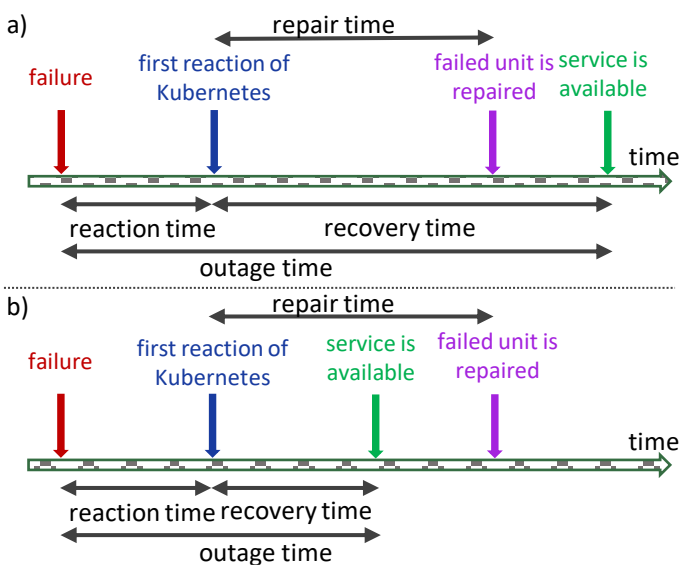

Fig. 6. Availability metrics. a) for the measurement of the architetcures without the SC. b) for the measurement of the architetcures with the SC.

**HA state assignment time:** The delay from the moment of sending the scaling request until the State Controller assigns the HA state to the last added pod.

In the following, the application is always scaled in or out by two. That is, the number of pod replicas is always even.

*C. Experiments, Results, and Analysis*

In this section, we evaluate the achievable availability as well as the scaling overhead of integrating the SC enriched with elasticity with Kubernetes. We aim at answering the research questions introduced earlier in the paper. Note that the measurements shown in the tables represent averages of 10 different experiments.

  *a) RQ1: What is the impact of the SC on the provided availability?*

To address this research question, we measure the availability metrics for the architectures in Fig. 1 and Fig. 2 ($n=1$) as a baseline. In these cases the number of pods will not impact the availability, because other pods would not know about the failure and do not have access to the state data of the failed pod, therefore cannot recover the service. We compare the results with those of the architectures in Fig. 4 and Fig 5 ($n=2$) where the SC is integrated with Kubernetes.

**Experiments**

In this set of experiments, we are interested in measuring the service outage when a failure happens. While our solution can handle different types of failures (e.g., pod process failure and node failure), we only consider the failure scenario where service outage is due to application container failure. In this scenario, the failure is simulated by killing the VLC container process from the OS. When Kubernetes detects the crash of the VLC container, it brings the pod to a state where the pod will not receive new requests. That is, it sets the pod's service state to "not ready". This time marks the reaction time. The repair time is marked by the time when Kubernetes restarts the VLC container and repairs the pod, i.e. its service state is "ready". The service is considered recovered when the video has started streaming from the last saved checkpoint before the failure.

**Results and Analysis**

The results of this set of experiments are shown in TABLE I. The results show a 46% improvement of the recovery time when the proposed SC is used with the StatefulSet controller and 55% improvement when the SC is used with the Deployment controller. The reason for this improvement is that, unlike in case of the architectures of Fig. 1 and Fig. 2, with the SC integrated we no longer need to wait for the failed pod to be repaired to have the service recovered. With the architectures of Fig 4 and Fig. 5, after the reaction of Kubernetes to the failure of the active pod (i.e. setting its service state "not ready"), the SC changes the "HAState" label and variable of the standby pod to active and accordingly this newly active pod will read the last stored state from the storage and resume the service. The results show that it takes longer for Kubernetes to repair the pod than it takes for the SC to assign the active HA state to the standby pod.

TABLE I. THE IMPACT OF INTEGRATING THE SC WITH KUBERNETES ON THE PROVIDED AVAILABILITY (RQ1).

| Architecture (unit = seconds) | Reaction Time | Repair Time | Recovery Time | Outage Time |
|---|---|---|---|---|
| StatefulSet controller (Fig. 2, $n=1$) | 0.679 | 1.029 | 1.480 | 2.159 |
| StatefulSet controller integrated with the SC (Fig. 4, $n=2$) | 0.719 | 1.083 | 0.793 | 1.512 |
| Deployment controller (Fig. 1, $n=1$) | 0.554 | 1.021 | 1.534 | 2.088 |
| Deployment controller integrated with the SC (Fig 5, $n=2$) | 0.784 | 1.244 | 0.688 | 1.472 |

We can also observe that integrating the SC and increasing the number of pods add some overhead to the reaction time. However, the increase in the reaction time is insignificant compared to the improvement in the recovery time.

  *b) RQ2: What is the impact of scaling during failover on the availability that the SC can provide for its managed microservices?*

To answer this question, we consider the architectures in Fig. 4 and Fig. 5 where the SC is integrated with Kubernetes. We measure the availability metrics when a scaling request is sent during the execution of a failover to evaluate the impact of simultaneous scaling and failure events.

**Experiments**

We conduct this set of experiments with two scenarios: scale-out and scale-in. For the scale-out scenario, we consider both architectures of Fig. 4 and Fig. 5 where we have one active and one standby pods ($n=2$). To simulate a failure, we forcefully kill the application container of the active pod, which streams the video. While the service is being recovered by the SC, we request to scale the application to four pods. We measure the availability metrics for the failed pod as well as the scaling time and the HA state assignment time for the added pods. We compare the availability metrics of this set of experiments with those where no scaling event has happened during failover. Moreover, we compare the scaling time and HA state assignment time of this set of experiments with those where no failure has happened during scaling the application.

In the scale-in scenario, we only consider the architecture with the StatefulSet controller and we have two active-standby pairs of pods ($n=4$). To simulate a failure we forcefully kill the application container of the "oldest" active pod which streams a video. To measure scaling time we also set the graceful termination period of pods to zero. meaning that when a pod is ordered to be terminated it is executed immediately. In this scenario, we measure the availability metrics for the failed pod as well as the scaling time for the deleted pod. We compare the availability metrics of these experiments with those where no scaling event had happened during failover. Again, we compare the scaling time of this set of experiments with those where no failure had happened during scaling the application.

## Results and Analysis

The results of this set of experiments are shown in TABLE II. The results show that when a scaling event happens during recovery, the outage time increases by 12% and 16%, respectively, for the scale-out and scale-in scenarios. We also evaluate the impact of scaling during failover on the scaling time by comparing them with the experiments where the only event is the scaling (without any simultaneous failure). The results of the experiments for both scale-out and scale-in scenarios show that when a failure happens during scaling, the scaling time increases by 66% for the scale-out and by 12% for the scale-in scenario. Moreover, for the scale-out scenario, the HA state assignment time also increases by 56% on average. The reason is that when scaling is triggered while a failover is in progress, the SC is busy with the failover process and it assigns the HA states with some delay.

It is important to note that TABLE II does not include results for the scale-in scenario for an architecture where the SC is used with a Deployment controller (as in Fig. 5). The reason is that in this case the scale-in request may result in deleting the standby pod of the failed active pod. Indeed, with this architecture, there is no order associated with the pods that would apply at their deletion, for example, when the application is scaled in. Therefore, it is not possible to guarantee the simultaneous execution of the service recovery and scale-in with this architecture.

*c) RQ3: What is the overhead of the SC at scaling?*

In this research question, we are interested in evaluating the impact of the SC on the time it takes for the application to be scaled by Kubernetes. We conduct the experiments with the architectures of Fig. 1, Fig. 2, Fig. 4, and Fig. 5.

**Experiments**

We conduct the experiments again for two scenarios: scale-out and scale-in. We again set the graceful termination period of the pods to zero. For the scale-out scenario, we consider all four architectures where the number of pods initially deployed is two ($n=2$). In each round of the experiments, we scale the application from two pods to $k$ pods where $k$ gets one of the values in {4, 8, 16, 32, 64, and 128}. For this scenario, we measure the scaling time and HA state assignment time.

For the scale-in scenario, we consider the same architectures. However, in each round of the experiments, the number of pods initially deployed (i.e., n) gets one of the values in {4, 8, 16, 32, 64, and 128}. In each round of the experiments, we scale in the application to 2 pods. For this scenario, we only measure the scaling time.

**Results and Analysis**

The measurements for the scale-out and scale-in scenarios are shown in TABLE III and TABLE IV, respectively. For the scale-out scenario (TABLE III), when the application is deployed as a StatefulSet, integrating the SC has a scaling overhead of 7.5% on average. Integrating the SC with a Deployment controller also increases the scaling time by 10.5% on average. The standard deviation for these measurements does not go above 23% of the average. Moreover, as it is shown in TABLE III and TABLE IV, the application deployed by a Deployment controller have shorter scaling and HA state assignment times compared to when it is deployed as a StatefulSet. The reason is that the pods deployed by Deployment controllers are created in parallel while with StatefulSet controllers they are created in sequence which takes more time. While fast start-up time can be considered as a benefit of deploying the applications with Deployment controllers, one should take into consideration that service protection is not guaranteed with Deployment controllers in scale-in scenarios. Indeed, Deployment controllers do not scale-in the application in an ordered manner and the active-standby pairs of pods might not be deleted together in the scale-in process. For the scale-in scenario (TABLE IV), the scaling overhead of the SC with StatefulSet controllers and with Deployment controllers is, respectively, 31% and 27% on average. The standard deviation for these measurements does not go above 28% of the average. Similar to the scale-out scenario, we also notice that applications deployed with Deployment controllers have a shorter scaling time. The reason is the same, i.e., that the pods deployed with Deployment controllers are deleted in parallel while with StatefulSet controllers, they are deleted in a predefined order which takes longer time.

TABLE II. EVALUATION OF THE PROVIDED AVAILABILITY WHEN SCALING HAPPENS DURING FAILOVER (RQ2).

| Scenario | Architecture | Event | Reaction Time | Repair Time | Recovery Time | Outage Time | Scaling Time | HA state Assignment time |
|---|---|---|---|---|---|---|---|---|
| Scale-out | StatefulSet controller integrated with the SC (Fig. 4, $n=2$) | Active pod fails | 0.719 | 1.083 | 0.793 | 1.512 | NA | NA |
| | | Application is scaled out to 4 | NA | NA | NA | NA | 4.234 | 5.653 |
| | | Scaling and failover overlap | 0.689 | 1.161 | 1.012 | 1.701 | 7.049 | 7.293 |
| | Deployment controller integrated with the SC (Fig. 5, $n=2$) | Active pod fails | 0.784 | 1.244 | 0.688 | 1.472 | NA | NA |
| | | Application is scaled out to 4 | NA | NA | NA | NA | 3.016 | 3.060 |
| | | Scaling and failover overlap | 0.607 | 1.205 | 1.028 | 1.635 | 5.055 | 5.608 |
| Scale-in | StatefulSet controller integrated with the SC (Fig. 4, $n=4$) | Active pod fails | 0.719 | 1.083 | 0.793 | 1.512 | NA | NA |
| | | Application is scaled in to 2 | NA | NA | NA | NA | 0.712 | NA |
| | | Scaling and failover overlap | 0.581 | 1.468 | 1.172 | 1.754 | 0797 | NA |

TABLE III. SCALING AND HA STATE ASSIGNMENT OVERHEAD FOR THE SCALE-OUT SCENARIO (RQ3).

| Architecture (*n*=2) | metric (unit: seconds) | 2 to 4 | 2 to 8 | 2 to 16 | 2 to 32 | 2 to 64 | 2 to 128 |
|---|---|---|---|---|---|---|---|
| StatefulSet controller (Fig. 2) | scaling time | 4.099 | 15.056 | 47.722 | 119.674 | 297.470 | 753.045 |
| StatefulSet controller integrated with the SC (Fig. 4) | scaling time | 4.234 | 16.692 | 49.937 | 131.726 | 312.037 | 842.068 |
| | HA state assignment time | 5.653 | 17.018 | 51.865 | 133.373 | 316.107 | 845.114 |
| Deployment controller (Fig. 1) | scaling time | 2.979 | 4.459 | 7.956 | 15.237 | 31.574 | 80.694 |
| Deployment controller integrated with the SC (Fig. 5) | scaling time | 3.016 | 4.914 | 8.856 | 17.428 | 35.248 | 92.240 |
| | HA state assignment time | 3.060 | 6.763 | 16.142 | 35.290 | 73.001 | 147.798 |

TABLE IV. SCALING OVERHEAD FOR THE SCALE-IN SCENARIO (RQ3).

| Architecture *n*={4, 8, 16, 32, 64, and 128} | metric (unit : seconds) | 4 to 2 | 8 to 2 | 16 to 2 | 32 to 2 | 64 to 2 | 128 to 2 |
|---|---|---|---|---|---|---|---|
| StatefulSet controller (Fig. 2) | scaling time | 0.555 | 1.353 | 2.613 | 5.459 | 11.440 | 26.062 |
| StatefulSet controller integrated with the SC (Fig. 4) | | 0.712 | 1.512 | 3.148 | 6.407 | 14.463 | 48.662 |
| Deployment controller (Fig. 1) | | 0.566 | 0.827 | 1.370 | 1.944 | 3.375 | 7.007 |
| Deployment controller integrated with the SC (Fig. 5) | | 0.641 | 1.327 | 1.555 | 2.375 | 4.441 | 8.821 |

*d) RQ4: What is the impact of simultaneous failures of active pods on the outage time of each failed pod?*

By this research question, we evaluate the SC performance in terms of availability when multiple active pods fail at the same time. Meaning that another failure happens while the SC is still in the process of handling the failover for the previously failed pod.

**Experiments**

We conduct a set of experiments with the architectures of Fig. 4 and Fig. 5 where the SC is used, respectively, with a StatefulSet controller and with a Deployment controller. For each architecture, the number of deployed pods is equal to ten ($n=10$). In each set of experiments, we simultaneously kill the application container of $k$ active pods where $k$ takes values from {1, 2, 3, 4, and 5}. In each round of the experiments, we measure the availability metrics for each failed pod separately and compare how simultaneous failures of multiple active pods affects their availability metrics.

**Results and Analysis**

The measurements of the experiments for RQ4 are shown in the diagrams of Fig. 7 and Fig. 8. These diagrams show that when multiple pods fail simultaneously, the later the pod's failure is detected, the longer it takes for the SC to recover the service for that pod. The reason is that the SC handles events in a FIFO (first-in-first-out) manner, so when a pod's failure is detected, it is put as an event in a queue. The SC will recover its service only after the recovery of other pods' that were inserted in the queue before this.

*D. Threats to Validity*

The following threats can affect the validity of our results. First, we conducted our experiments in a relatively small Kubernetes cluster. In larger clusters where the number of nodes is high, Kubernetes may behave differently which might result in different availability and scaling overhead measurements as the Kubernetes' performance can change in such clusters. Also, for the scaling overhead measurements, the maximum number of pods that could be deployed in a reasonable time was limited to around 128 pods for a cluster with 8 worker nodes. With larger clusters, it will be possible to measure the scaling time for a larger number of pods and reach more accurate results.

Another threat to the validity of our results is related to the tools and mechanisms used in our experiments for measuring the time. We rely on the timestamps reported in Kubernetes and Docker logs. However, one can instrument the containers in order to achieve a more precise measurement. Finally, we only considered the case of an on demand video streaming application and other types of applications should be considered before generalizing the results.

## VI. RELATED WORK

Researchers and practitioners have adopted the microservices paradigm in several domains, such as the cloud computing [20, 21, 22], service computing [23, 24], internet of things [25], to take advantage of its benefits both in the development and operational phases [20, 26, 27]. In this paper, we are interested in the operational phase. Thus, in our review of related work, we first discus the state of the art on microservices and containers orchestration, before focusing on works related to the availability of stateful microservice-based applications. Table V summarizes this section by categrozing all related work, including our paper, with respect to certain criteria such as workload type and management objectives.

Kubernetes is the de-facto open-source container orchestration platform. Several recent studies built on Kubernetes to propose enhancement and more efficient container scheduling and orchestration approaches. In [28] Zhong et al. propose a task allocation strategy to make container scheduling and scaling decisions in a cost-efficient manner through resource utilization optimization and elastic resource pricing. Three main features were considered, first the support of heterogeneous job configurations to optimize the initial placement of containers into existing resources by task packing. Second, a cluster size adjustment to meet the changing workload through autoscaling algorithms is proposed. Finally, the rescheduling mechanism to shut down underutilized VM instances for cost-saving and reallocate the relevant jobs without losing task progress is considered. Pascinsk et al. [29] developed a Kubernetes-based Global Cluster Manager specialized in geographic orchestration of network-intensive workloads. It supports autonomic task arrangement. This manager automatically selects the best geographically available computing resource within the defined data centres according to a QoS model of the software components. Chung et al. [30] proposed Stratus a container-based cluster scheduler designed for batch job scheduling on public clouds. It exploits the cloud properties and runtime estimates to reduce the cost of cluster jobs execution by packing jobs that should complete around the same time. Stratus aggressively packs tasks into machines, it uses task migration to clear under-utilized instances. Stratus actively avoids having leased machines that are not highly utilized, trying to make allocated resources be either mostly full or empty, so they can be released to reduce the cost. In [31], the authors designed a customized scheduler on top of the Kubernetes platform by extending the existing rescheduling feature for better arrangements of task co-location. The proposed solution integrates the use of schedulers, autoscalers, and reschedulers as a mechanism to make container orchestration systems cloud-aware. The scheduler optimizes the initial placement of containers, the autoscaler enables the current demand for resources to be met and underutilized or idle nodes to be shut down, while the rescheduler allows for the initial placement of containers to be revised at runtime for better resource utilization. Similarly, to the previously mentioned works, our solution builds on Kubernetes. However, in contrast to the aforementioned works, we rely on the native scheduling and orchestration mechanisms of Kubernetes. We provide an availability management solution without altering the default behavior of Kubernetes. This will ease the integration and the adoption of the availability management solution.

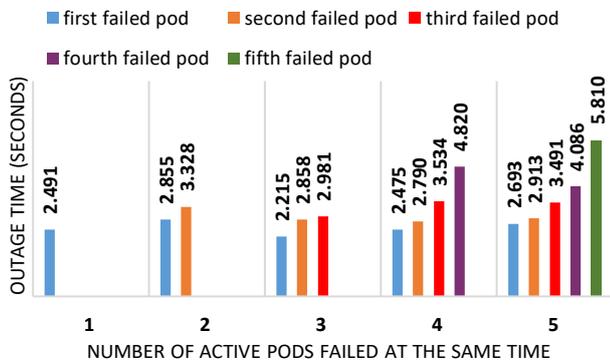

Fig. 7. Service Outage of simultaneously failed pods – The SC integrated with a StatefulSet controller (RQ4).

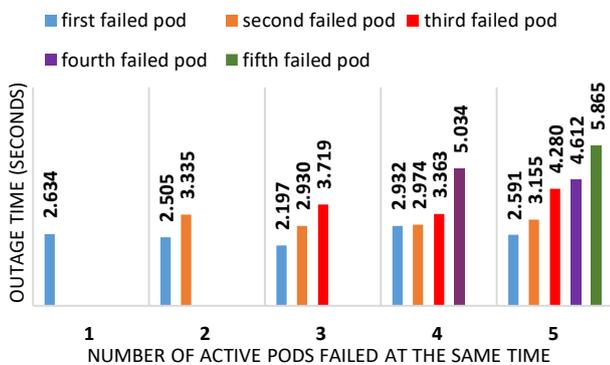

Fig. 8. Service Outage of simultaneously failed pods – The SC integrated with a Deployment controller (RQ4).

Kang et al. in [17] propose a microservice based architecture and use containers to operate and manage the cloud infrastructure services. In their architecture, each container is monitored by a sidekick container and in case of failure, recovery actions are taken. In their experiments, the considered stateful microservice is a MySQL database in the active-active mode. For synchronizing data between microservice instances, they suggest shared storage and application level data replication. In the former, all MySQL microservice instances access the same data while in the latter the database process replicates the data across the cluster. However, one cannot guarantee service recovery and continuity only by replicating the state data between microservice instances. For recovery with service continuity, a microservice instance needs to detect or be notified about the failure to access the replicated data and continue the service. Also, it should be clear where the state data of each microservice instance is stored. In our solution, a state replication mechanism is provided through which the active pod can replicate its state data to the standby pod and a third party (the HA State Controller) notifies the standby pod if its corresponding active pod fails. Moreover, each pod stores its data separately and is aware of the location of its data. From the perspective of the adoption of the proposed solution in practical scenarios, the authors in [17] opted for their own implementation of a microservice architecture instead of building on top of the existing ecosystem such as Kubernetes and Docker Swarm. Our solution is based on Kubernetes' principles and has been integrated easily with Kubernetes, and thus it could easily be adopted in practice. Our solution closes existing gaps in Kubernetes with respect to stateful microservice-based applications.

Netto et al. in [18] propose KRaft, an incorporation of the Raft [19] consensus algorithm in Kubernetes for state machine replication. With this incorporation, requests sent to the Kubernetes managed application can be executed in the same order by all containers which results in synchronizing the states of all containers. In their work, containers include processes for communicating with the Kubernetes API server. Through this communication, each container periodically asks about the IP addresses of other replicas. In the original Raft algorithm, it is only the leader that can accept requests. However, in Kubernetes, all replicas can receive requests. Therefore, in their work, they made modifications to the containers so when a non-leader replica receives a request, it redirects the request to the leader and after the request is executed, the leader receives the request and sends it back to the client. This way, if the leader fails, other replicas can become the leader and continue the service. Compared to our solution, the level of protection provided by the KRaft algorithm is stronger. However, this comes at the cost of increase in the resource usage. With the KRaft algorithm, to handle one failure, for each container that provides a service two extra containers need to be running. With our solution, however, one extra container is enough for recovering the service after a single failure. With respect to the methodology, the solution in [18] was only evaluated from the perspective of resource consumption (CPU, memory, throughput, and latency). However, a key characteristics of microservices and application managed with Kubernetes is their flexibility of scaling. This has not been handled in [18] in contrast to our solution and the experiments we conducted.

TABLE V. COMPARISON WITH RELATED WORK

| Related work | Supported workload type | | Container management objective | | |
|---|---|---|---|---|---|
| | Stateful | Stateless | Elasticity | Availability | Scheduling |
| [17] | X | | | X | |
| [18] | X | | | X | |
| [28] | | X | X | | X |
| [29] | | X | | | X |
| [30] | | X | X | | X |
| [31] | | X | | | X |
| Our work | X | | X | X | |

VII. CONCLUSION

In this paper, we evaluated the availability provided by Kubernetes' repair actions for stateful applications and proposed a solution to improve this availability. Our solution enables the recovery of the service independently from the repair of the failed pod. It proposes a SC which allows for failure handling at the platform (i.e., Kubernetes) level by automatically redirectig services to healthy pods through the management of secondary labels. The SC is also capable of managing the availabilty of the application when it is scaled, i.e. the number of pods changes, by forming and maintaining pairs of active-standby pods.

In our evaluations of the SC it was shown that integrating our solution improved service recovery by 50% on average. However, we observed that when a scaling event happened while the SC was carrying out a failover, the outage time increased by 16% and the HA state assignment time also increased by 48% for the scale-out scenario. Moreover, we measured the scaling overhead of the SC integrated with Kubernetes between 7.5% and 10.5%. We also observed that both scaling and HA state assignment were done faster when the application was deployed by a Deployment controller as opposed to when it is deployed by a StatefulSet controller. Indeed, unlike StatefulSet controllers, the Deployment controllers do not add or delete pods in any particular order and one by one. While the fast deployment and HA state assignment of pods can be considered as a reason to deploy an application by a Deployment controller, one should consider the potential drawbacks of deploying stateful applications with Deployment controllers as well. In particular, considering our SC with Deployment controllers, service protection cannot be guaranteed in and after scale-in, because Deployment controllers do not scale-in the application pods in a guaranteed order and therefore active-standby pairs of pods might not be deleted at scale-in together. We also evaluated the availability provided by the SC when multiple active pods failed simultaneously, and observed that the later the failure of a pod was detected by the SC, the longer its recovery took.

We identify high resource usage as a limitation of our solution which is related to the 2N redundancy model where each standby microservice protects only one active microservice. As future work, this limitation can be addressed by implementing other redundancy models to share a standby microservice instance between a number of active microservice instances. Another limitation of our solution is that service protection is not guaranteed with Deployment controllers in scale-in scenarios. Our solution can be modified to reassign active and standby states to remaining pods after a scale-in so we can ensure that no active (or standby) pod will lose its peer.

ACKNOWLEDGMENT

This work has been partially supported by Natural Sciences and Engineering Research Council of Canada (NSERC) and Ericsson.